\documentstyle[twoside]{article}

\catcode`\@=11
\long\def\@makefntext#1{
\protect\noindent \hbox to 3.2pt {\hskip-.9pt  
$^{{\eightrm\@thefnmark}}$\hfil}#1\hfill}		

\def\@makefnmark{\hbox to 0pt{$^{\@thefnmark}$\hss}}	
	
\def\ps@myheadings{\let\@mkboth\@gobbletwo
\def\@oddhead{\hbox{}
\rightmark\hfil\eightrm\thepage}   
\def\@oddfoot{}\def\@evenhead{\eightrm\thepage\hfil
\leftmark\hbox{}}\def\@evenfoot{}
\def\sectionmark##1{}\def\subsectionmark##1{}}

\oddsidemargin=\evensidemargin
\newcounter{sectionc}\newcounter{subsectionc}\newcounter{subsubsectionc}
\renewcommand{\section}[1] {\vspace{12pt}\addtocounter{sectionc}{1} 
\setcounter{subsectionc}{0}\setcounter{subsubsectionc}{0}\noindent 
	{\tenbf\thesectionc. #1}\par\vspace{5pt}}
\renewcommand{\subsection}[1] {\vspace{12pt}\addtocounter{subsectionc}{1} 
	\setcounter{subsubsectionc}{0}\noindent 
	{\bf\thesectionc.\thesubsectionc. {\kern1pt \bfit #1}}\par\vspace{5pt}}
\renewcommand{\subsubsection}[1] {\vspace{12pt}\addtocounter{subsubsectionc}{1}
	\noindent{\tenrm\thesectionc.\thesubsectionc.\thesubsubsectionc.
	{\kern1pt \tenit #1}}\par\vspace{5pt}}
\newcommand{\nonumsection}[1] {\vspace{12pt}\noindent{\tenbf #1}
	\par\vspace{5pt}}

\newcommand{\textlineskip}{\baselineskip=13pt}

\def\eightcirc{
\begin{picture}(0,0)
\put(4.4,1.8){\circle{6.5}}
\end{picture}}
\def\eightcopyright{\eightcirc\kern2.7pt\hbox{\eightrm c}} 

\def\abstracts#1#2#3{{
	\centering{\begin{minipage}{4.5in}\baselineskip=10pt\footnotesize
	\parindent=0pt #1\par 
	\parindent=15pt #2\par
	\parindent=15pt #3
	\end{minipage}}\par}} 

\renewenvironment{thebibliography}[1]
	{\frenchspacing
	 \ninerm\baselineskip=11pt
	 \begin{list}{\arabic{enumi}.}
        {\usecounter{enumi}\setlength{\parsep}{0pt}     
	 \setlength{\leftmargin 12.7pt}{\rightmargin 0pt} 
         \setlength{\itemsep}{0pt} \settowidth
	{\labelwidth}{#1.}\sloppy}}{\end{list}}

\newcounter{itemlistc}
\newcounter{romanlistc}
\newcounter{alphlistc}
\newcounter{arabiclistc}

\def\@citex[#1]#2{\if@filesw\immediate\write\@auxout
	{\string\citation{#2}}\fi
\def\@citea{}\@cite{\@for\@citeb:=#2\do
	{\@citea\def\@citea{,}\@ifundefined
	{b@\@citeb}{{\bf ?}\@warning
	{Citation `\@citeb' on page \thepage \space undefined}}
	{\csname b@\@citeb\endcsname}}}{#1}}

\newif\if@cghi
\def\cite{\@cghitrue\@ifnextchar [{\@tempswatrue
	\@citex}{\@tempswafalse\@citex[]}}
\def\citelow{\@cghifalse\@ifnextchar [{\@tempswatrue
	\@citex}{\@tempswafalse\@citex[]}}
\def\@cite#1#2{{$\null^{#1}$\if@tempswa\typeout
	{IJCGA warning: optional citation argument 
	ignored: `#2'} \fi}}

\def\@refcitex[#1]#2{\if@filesw\immediate\write\@auxout
	{\string\citation{#2}}\fi
\def\@citea{}\@refcite{\@for\@citeb:=#2\do
	{\@citea\def\@citea{, }\@ifundefined
	{b@\@citeb}{{\bf ?}\@warning
	{Citation `\@citeb' on page \thepage \space undefined}}
	\hbox{\csname b@\@citeb\endcsname}}}{#1}}

\def\@refcite#1#2{{#1\if@tempswa\typeout
        {IJCGA warning: optional citation argument
	ignored: `#2'} \fi}}

\def\refcite{\@ifnextchar[{\@tempswatrue
	\@refcitex}{\@tempswafalse\@refcitex[]}}

\def\pmb#1{\setbox0=\hbox{#1}
	\kern-.025em\copy0\kern-\wd0
	\kern.05em\copy0\kern-\wd0
	\kern-.025em\raise.0433em\box0}

\def\fnt#1#2{\footnotetext{\kern-.3em
	{$^{\mbox{\scriptsize #1}}$}{#2}}}

\def\runninghead#1#2{\pagestyle{myheadings}
\markboth{{\protect\footnotesize\it{\quad #1}}\hfill}
{\hfill{\protect\footnotesize\it{#2\quad}}}}
\font\tenrm=cmr10
\font\tenit=cmti10 
\font\tenbf=cmbx10
\font\bfit=cmbxti10 at 10pt
\font\ninerm=cmr9

\font\eightrm=cmr8

\def\qed{\hbox{${\vcenter{\vbox{			
   \hrule height 0.4pt\hbox{\vrule width 0.4pt height 6pt
   \kern5pt\vrule width 0.4pt}\hrule height 0.4pt}}}$}}

\begin{document}

\runninghead{H.C. Rosu, Stationary VFN
$\ldots$} {H.C. Rosu, Nonstationary VFN
$\ldots$}


\normalsize\textlineskip
\thispagestyle{empty}
\setcounter{page}{1}


\vspace*{0.88truein}

\centerline{gr-qc/9912056 v2}
\vspace*{0.035truein}
\centerline{\bf STATIONARY AND NONSTATIONARY SCALAR VACUUM FIELD NOISES}
\vspace*{0.035truein}
\vspace*{0.37truein}
\centerline{\footnotesize HARET C. ROSU
\footnote{e-mail: rosu@ifug3.ugto.mx}
}
\vspace*{0.015truein}
\centerline{\footnotesize\it Instituto de F\'{\i}sica,
Universidad de Guanajuato, Apdo Postal E-143, Le\'on, Gto,
Mexico}
\baselineskip=10pt
\vspace*{10pt}
\vspace*{0.225truein}


\vspace*{0.13truein}
\abstracts{{\bf Summary}.-
If stationary, the spectrum of vacuum field
noise (VFN) is an important ingredient to get
information about the curvature
invariants of classical worldlines (relativistic classical trajectories).
For scalar quantum field vacua there are six stationary
cases as shown by Letaw some time ago, these are reviewed here. However, the
non-stationary vacuum noises are not out of reach and can be processed by a
few mathematical methods which I briefly comment on. Since the information
about the kinematical curvature invariants of the worldlines is of radiometric
origin, hints are given on a more useful application to radiation and beam 
radiometric standards at relativistic energies.
\\
}{}{}

\section{Introduction}

\noindent
Soon after Hawking's theoretical discovery of black hole evaporation,
a number of authors have used scalar relativistic quantum field theory in
order to show that apparently there exist thermal radiation effects due
entirely to
vacuum fluctuations in the case of relativistic noninertial motion.
In particular, Unruh [\refcite{unr}] studied a
uniformly linearly accelerated two-level detector of DeWitt (DW)
monopole type moving in flat space
and proved that for it the Minkovski vacuum looks like a heat bath at a
temperature $T_{U}=a/2\pi$ ($\hbar=c=1$), where $a$ is the proper
acceleration. Although not a realistic result, it
is simple
and general, and connected to Hawking's temperature that may be considered
the case in which $a$ is the gravitational acceleration at
the Schwarzschild horizon.
By 1986, Takagi wrote an excellent review where he used the concept
of vacuum noise for the excitations of the vacuum as seen in noninertial
relativistic motion [\refcite{taka}].
However, for more general trajectories
two or all three kinematical invariants of Frenet type get involved, not just
the curvature one as in
the case of linear acceleration. A more detailed study of scalar vacuum noise,
with significant results has been accomplished by Letaw, either alone or in
collaboration with Pfautsch [\refcite{let}].
My aims here are (i) to review Letaw's results, that I present in a different
perspective in which I emphasize the radiometric features of the
curvature invariants of stationary worldlines and (ii) to provide a few hints
on the methods that may be used to analyze the nonstationary cases.

\section{Six types of stationary scalar VFNs}  

\noindent
In general, the scalar quantum field vacua
do not possess stationary
vacuum excitation spectra (abbreviated as SVES)
for all types of classical relativistic trajectories on which the DW detector
moves. Nevertheless, the linear uniform
acceleration is {\em not} the only case with that property as was shown by
Letaw who
extended Unruh's considerations, obtaining six types
of worldlines with SVES for DW monopole detectors
(SVES-1 to SVES-6, see below). The line of arguments is the following.
The DW detector is effectively immersed in a scalar bath. Its rate of
excitation is determined by the energy
spectrum of the scalar bath that can be expressed as the density of states
times a cosine Fourier transform of the Wightman correlation
function (WCF) of the scalar field. Since the WCF is directly expressed in
terms of the inverse of the geodetic interval what one needs
to calculate is a Fourier transform of the inverse of the geodetic
interval $dx_{\mu}^2$. Some calculations are sketched in the Appendix.
For stationarity, one should assume the WCF as dependent only on the proper
time interval.
As shown by Letaw, the stationary worldlines
are solutions of some generalized Frenet equations
on which the condition of constant curvature invariants is imposed,
i.e., constant curvature $\kappa$, torsion $\tau$, and
hypertorsion $\nu$, respectively. Notice that one can employ other frames
such as the Newman-Penrose spinor formalism as recently
did Unruh [\refcite{nepe}] but the Frenet-Serret one is in overwhelming use
throughout physics. The six stationary cases are the following

\bigskip
\underline{1. $\kappa =\tau=\nu=0$},
(inertial, uncurved worldlines). SVES-1 is a trivial cubic spectrum
\begin{equation}
S_1(E)=\frac{E^3}{4\pi ^2}~,
\end{equation}
i.e., as given by a vacuum of zero point energy per mode $E/2$
and density of states $E^2/2\pi ^2$.

\bigskip
\underline{2. $\kappa \neq 0$, $\tau=\nu=0$},
(hyperbolic worldlines). SVES-2 is Planckian allowing the
interpretation of $\kappa/2\pi$ as `thermodynamic' temperature. In the
dimensionless variable $\epsilon _{\kappa}=E/\kappa$ the vacuum spectrum reads
\begin{equation}
S_2(\epsilon _{\kappa})
=\frac{\epsilon _{\kappa}^{3}}{2\pi ^2(e^{2\pi\epsilon _{\kappa}}-1)}~.
\end{equation}

\bigskip
\underline{3. $|\kappa|<|\tau|$, $\nu=0$, $\rho ^2=\tau ^2-\kappa ^2$},
(helical worldlines). SVES-3 is an analytic function
corresponding to the case 4 below only in the limit $\kappa\gg \rho$
\begin{equation}
S_3(\epsilon _{\rho})\stackrel{\kappa/\rho\rightarrow \infty}
{\longrightarrow} S_4(\epsilon _{\kappa})~.
\end{equation}
Letaw plotted the numerical integral $S_3(\epsilon _{\rho})$,
where $\epsilon _{\rho}=E/\rho$ for various values of $\kappa/\rho$.

\bigskip
\underline{4. $\kappa=\tau$, $\nu=0$},
(the spatially projected worldlines are the semicubical parabolas
$y=\frac{\sqrt{2}}{3}\kappa x^{3/2}$ containing a
cusp where the direction of motion is reversed). SVES-4 is analytic, and
since there are two equal curvature invariants one can use the
dimensionless energy variable $\epsilon _{\kappa}$
\begin{equation}
S_{4}(\epsilon _{\kappa})= \frac{\epsilon _{\kappa}^{2}}{8\pi ^2 \sqrt{3}}
e^{-2\sqrt{3}\epsilon _{\kappa}}~.
\end{equation}
It is worth noting that $S_4$, being a monomial times an exponential,
is rather close to the Wien-type spectrum
$S_{W}\propto\epsilon ^3e^{- {\rm const.}\epsilon}$.

\bigskip
\underline{5.  $|\kappa|>|\tau|$, $\nu=0$, $\sigma ^2=\kappa ^2-\tau ^2$},
(the spatially projected worldlines are catenaries, i.e., curves of the type
$x=\kappa \cosh (y/\tau)$). In general, SVES-5 cannot be found
analytically. It is an intermediate case, which
for $\tau/\sigma\rightarrow 0$ tends to SVES-2,
whereas for $\tau/\sigma\rightarrow\infty$ tends toward SVES-4
\begin{equation}
S_2(\epsilon _{\kappa})
\stackrel{0\leftarrow \tau/\sigma}{\longleftarrow}
S_5(\epsilon _{\sigma})\stackrel{\tau/\sigma\rightarrow \infty}
{\longrightarrow}S_4(\epsilon _{\kappa})~.
\end{equation}

\bigskip
\underline{6. $\nu\neq 0$},
(rotating worldlines uniformly accelerated normal to their plane of rotation).
According to Letaw,
SVES-6 forms a two-parameter set of curves. These trajectories are
a superposition of the constant linearly accelerated motion and uniform
circular motion. The corresponding vacuum spectra have not been calculated
by Letaw, not even numerically.

Thus, only the hyperbolic worldlines, having just one nonzero curvature
invariant, allow for a Planckian SVES and for a strictly one-to-one
mapping between the curvature invariant $\kappa$ and the `thermodynamic'
temperature ($T_{U}=\kappa /2\pi$). The VFN of semicubical
parabolas can be fitted by Wien-type spectra, the radiometric parameter
corresponding to both curvature and torsion.
The other stationary cases, being nonanalytical, lead to
approximate determination of the curvature invariants defining locally the
classical worldline on which a relativistic quantum particle moves.

\section{Preferred vacua and/or high energy radiometric standards}

\noindent
There is much interest in considering the
magnetobremsstrahlung radiation patterns at
accelerators in the aforementioned perspective [\refcite{mont}]. The `thermal
quantum field vacuum' standpoint has been initiated by
the works of Bell and collaborators during 1984-1987 [\refcite{bell}]
In this sense, a sufficiently general and acceptable statement
on the {\em universal} nature of
the kinematical Frenet parameters occuring in a few important
quantum field model problems can be formulated as follows

\bigskip

\noindent
{\em  There exist accelerating
classical trajectories (worldlines) on which moving ideal (two-level)
quantum systems can detect the scalar vacuum environment as a stationary
quantum field vacuum noise with a spectrum directly related to
the curvature invariants of the worldline, thus allowing for a
radiometric meaning of those invariants}.

\bigskip

\noindent
Although this may look an extremely ideal (unrealistic) formulation
for accelerator radiometry [\refcite{arad}],
where the spectral photon flux formula of
Schwinger,\cite{sflux} is very effective (synchrotron flux 
$\propto \xi \int _{\xi}^{\infty}
K_{5/3}(z)dz$, with the variable $\xi$ scaled in terms of the
radian frequency at maximum power),
I recall that Hacyan and Sarmiento [\refcite{hs}] developed a formalism
similar to the scalar case to calculate the vacuum stress-energy tensor of the
electromagnetic field in an arbitrarily moving frame and applied it to a
system in uniform rotation, providing formulas for the energy density,
Poynting flux and stress of zero-point oscillations in such a frame.
Moreover, Mane [\refcite{mane}] has suggested the Poynting flux of Hacyan and
Sarmiento to be in fact synchrotron radiation when it is coupled to an
electron.

\noindent
Another important byproduct, that I put forth
in a previous work [\refcite{rin}], is the possibility to choose a class of
preferred vacua of the quantum world [\refcite{pr}] as {\em all} those
having stationary vacuum noises with
respect to the classical (geometric) worldlines of {\em constant} curvature
invariants because in this case one may find some
necessary attributes of universality in the more general
quantum field radiometric sense [\refcite{rad}], in which the Planckian Unruh
thermal spectrum is included as a particularly important case.
Of course, much work remains to be done for a more ``experimental" (i.e.,
realistic)
picture of highly academic calculations in quantum field theory,
but a careful look to the literature
shows that there are already definite steps in this direction
[\refcite{steps}].
Even Unruh, in one of his last studies on the `orbiting electron', had
to conclude that it does respond to a `thermal' bath, but one with a
frequency-dependent temperature [\refcite{ur}].
One should notice that the heat-bath picture and the associated VFNs look
extremely ideal from the
experimental standpoint and not very convincing with respect with the more
definite and well-settled quantum electrodynamical calculations.
Indeed, it is known that only strong external fields
can make the quantum electrodynamical vacuum to react and show its physical
properties, becoming similar to a magnetized and polarized medium,
and only by such means one can learn about the nature of the instabilities
and the physical structure of the QED vacuum.
For important results regarding the relationship between Schwinger mechanism
and Unruh effect, the reader is directed to some recent works [\refcite{gab}].
I also would like to point out that the VFN stochastic processes are directly
related to the motion of single particles that would be difficult to
disentangle within the bunches of particles circulating in storage rings.
The bunch motion is the result of more common and powerful sources able to
produce various
stochasticity features of the beam [\refcite{apaul}]. There were proposals
to detect VFNs in somewhat cleaner environments, such as
Penning traps of {\em geonium} type, in which the motion of a single
electron can be monitored for a long time [\refcite{geon}].

\section{Nonstationary VFNs}

\noindent
The nonstationary VFNs
have a time-dependent spectral content requiring joint
time and frequency information, i.e. generalizations of the power spectrum
analysis such as tomographical processing [\refcite{manko}] and wavelet
transform analysis [\refcite{wavl}].
The so-called non-commutative tomography (NCT) transform
$M(s,\mu,\nu)$ proposed by Manko and Vilela Mendes seems to be an attractive
way of processing the analytic nonstationary signals and perhaps one can
hope to get unambiguous information for nonanalytic signals as well.
In the definition of $M$, $s$ is just an arbitrary curve in the
non-commutative time-frequency plane. The most simple examples have been
given by Manko and Vilela Mendes by means of the axes $s=\mu t+\nu \omega$,
where $\mu$ and $\nu$ are linear combination parameters.
The NCT transform is related to the Wigner-Ville
quasidistribution $W(t,\omega)$ by an invertible transformation of the form
\begin{equation}
M(s,\mu,\nu)=\int\exp[-ik(s-\mu t- \nu \omega)]W(t,\omega)\frac{dkd\omega dt}
{(2\pi)^2}~.
\end{equation}
According to Man'ko and Vilela Mendes the well-known interpretation ambiguities
of the Wigner-Ville transform can be avoided in the case of
NCT transform, i.e., a rigorous probability interpretation can be assigned.
Other useful properties are $M(t,1,0)=|f(t)|^2$ and $M(\omega,0,1)=
|f(\omega)|^2$, where $f$ is the analytic signal which is simulated by $M$.
But perhaps the most interesting property of $M$ is the possibility to
detect the presence of signals in noises for small signal-to-noise ratios.
This last property may be very effective for detecting VFNs which
are very small `signals' with respect to more common noises.

On the other hand,
since in the quantum detector method the vacuum autocorrelation functions
are the essential physical quantities, and since according to
fluctuation-dissipation theorem(s) (FDT) they are related to the linear
(equilibrium) response functions to an initial condition/vacuum,
the FDT approach has been of interest for years [\refcite{hutera}].
Here, I would like to point out the recent
generalization of FDT to some classes of out of
equilibrium relaxational systems [\refcite{out}] that looks promising for the
nonstationary ``vacuum baths" as well. At the formal level, the generalization
is quite simple by introducing a two-time FD ratio $X(t,t^{'})$
in the FDT, i.e.
one writes $T_{\rm eff}R(t,t^{'})=X(t,t^{'})
\frac{\partial C}{\partial t^{'}}(t,t^{'})$,
where $T_{\rm eff}$ is an effective temperature, $R$ is the response function,
and $C$ is the associated correlation function. It is the FD ratio that is
employed to perform the separation of scales.
There are already interesting results regarding out of equilibrium
effective temperatures, which in general are time-scale-dependent
quantities [\refcite{eft}].
Precisely such kind of quantities are needed for
relativistic VFNs [\refcite{ur}],
which correspond naturally to out of equilibrium conditions.

\section{Conclusion}

The main conclusion of this work is that one may well ascribe radiometric
meaning to the curvature invariants of stationary worldlines. This means
that these invariants may be used as radiometric quantities beyond the
common ``thermodynamic" temperature, which is of limited value in the
worldline approach.
For nonstationary VFNs, the usage of NCT transform and/or considerations of
FDT type may be quite useful in the processing of the VFN signals.
The experimental challenge is of course the detection of VFN spectra.

\bigskip

\noindent
{\bf Acknowledgment}

\bigskip

\noindent
This work was partially supported by the CONACyT
project 458100-5-25844E.

\bigskip
\setcounter{equation} {0}

\noindent
\appendix{\bf Appendix: Getting the SVES formulas}

\bigskip

\noindent
One can calculate all sorts of SVES by means of the following general formula
$$
S(E)=|\frac{E^2}{4\pi ^3}\int _{-\infty}^{\infty}e^{-iEs}\left(
[x_{\mu}(s)-x_{\mu}(0)][x^{\mu}(s)-x^{\mu}(0)]\right)^{-1}ds|=
\frac{E^2}{4\pi ^3}|{\rm I}|~,
\eqno(A1)
$$
where $x^{\mu}(s)$ is an arbitrary point on the worldline and $x^{\mu}(0)$ is
the initial point. The signature of the Minkowski
metric is $\eta _{\mu \nu}=(1,-1,-1,-1)$.
I confirm Letaw's results by sketching the calculation of the
integral ${\rm I}$ for the six stationary cases.
Simple details that have been skipped by Letaw can be found here.
In general, the VFN spectra present a fourth power of a Frenet-type
invariant as a scaling parameter ($\kappa$, $\rho$ and $\sigma$, respectively)
that should be taken into account when
performing the calibration of the spectra.

\bigskip

\noindent
1. {\bf The recta}.

\noindent
$x^{\mu}(s)=(s,0,0,0)$; $x^{\mu}(0)=(0,0,0,0)$. The integral is
$$
{\rm I}_1= \int _{-\infty}^{+\infty} \frac{e^{-iEs}}{s^2}ds~.
\eqno(A2)
$$
It can be obtained by (i) Cauchy's residue theorem and
(ii) $e^{-iEs}=(1-iEs+...)$.
The value of the integral is
$\pi i(-iE)=\pi E$, and therefore one gets the cubic spectrum.
This inertial zero-point cubic spectrum will appear in all the other five
stationary spectra as an
additive background and therefore one may take into account only the
non-cubic contributions as a measure of noninertial vacuum effects.

\bigskip
\noindent
2. {\bf The hyperbola}.

\noindent
$x^{\mu}(s)=\kappa ^{-1}(\sinh \kappa s,\cosh \kappa s,0,0)$ and
$x^{\mu}(0)=\kappa ^{-1}(0,1,0,0)$. Now the
integral is
$$
{\rm I}_2=\int _{-\infty}^{+\infty}\frac{e^{-i\epsilon _{\kappa}u}}{2(\cosh u
-1)}du~.
\eqno(A3)
$$
Writing $e^{-i\epsilon _{\kappa}u}=\cos\epsilon _{\kappa} u -
i\sin\epsilon _{\kappa} u$, one makes use of formula 3.983.3 at page 505
in the fourth edition of the {\em Table} of
Gradshteyn and Ryzhik (GR) to get
$$
\int _{0}^{+\infty}\frac{{\rm cos}ax}{\cosh x -1}dx=-(\pi a){\rm cth}(\pi a)~,
\eqno(A4)
$$
and of the Cauchy theorem for the integral in the sinus function
$$
-i\int _{-\infty}^{+\infty}
\frac{{\rm sin}(\epsilon _{\kappa}x)dx}{{\rm cosh} x -1}=
\pi \epsilon _{\kappa}~.
\eqno(A5)
$$
Thus
$$
{\rm I}_2=-\pi\epsilon _{\kappa}{\rm cth}(\pi\epsilon _{\kappa})
+\frac{\pi \epsilon _{\kappa}}{2}=\pi\epsilon _{\kappa}[1-{\rm cth}
(\pi\epsilon _{\kappa})]-\frac{\pi \epsilon _{\kappa}}{2}=
-2\pi\epsilon _{\kappa}\frac{1}{e^{2\pi \epsilon _{\kappa}}-1}
-\frac{\pi\epsilon _{\kappa}}{2}~,
\eqno(A6)
$$
where the first term leads to the Planckian spectrum and
the latter to the cubic zero-point contribution.

\bigskip
\noindent
3. {\bf The helix}.

\noindent
$x^{\mu}(s)=\rho ^{-2}(\tau\rho s,\kappa
{\rm cos}\rho s,\kappa\sin \rho s,0)$, $x^{\mu}(0)=\rho ^{-2}(0,\kappa,0,0)$.
The integral reads
$$
{\rm I}_3=\rho ^{3}\int _{-\infty}^{+\infty}
\frac{e^{-i\epsilon _{\rho}u}}
{2\kappa ^{2}({\rm cos}u-1)+\tau ^{2}u^2}du~. 
\eqno(A7)
$$
According to Letaw this integral is non-analytic and indeed I was not able
to find any helpful formula in the GR {\em Table}.

\bigskip
\noindent
4. {\bf The semicubical parabola}.

\noindent
$x^{\mu}(s)=(s+\frac{1}{6}\kappa ^2s^3, \frac{1}{2}\kappa s^2,
\frac{1}{6}\kappa ^2s^3,0)$, $x^{\mu}(0)=(0,0,0,0)$. The integral reads
$$
{\rm I}_4
=\kappa\int _{-\infty}^{+\infty}\frac{e^{-i\epsilon _{\kappa}u}du}{u^2(1
+\frac{1}{12}u^2)}=
\kappa{\rm I}_1-\kappa\int _{-\infty}^{+\infty}
\frac{e^{-i\epsilon _{\kappa}u}du}{(12+u^2)}~.
\eqno(A8)
$$
Of interest is only the second integral that can be found in the
GR {\em Table} at page 359
$$
\int _{-\infty}^{+\infty}\frac{e^{-ipx}dx}{a^2+x^2}=\frac{\pi}{|a|}e^{-|ap|}~,
\eqno(A9)
$$
for $a>0$ and $p$ real. Thus, one gets
$$
\int _{-\infty}^{+\infty}\frac{e^{-i\epsilon _{\kappa}u}du}{(12+u^2)}=
\frac{\pi}{\sqrt{12}}e^{-\sqrt{12}\epsilon _{\kappa}}~.
\eqno(A10)
$$
The final result is
$$
S_4=
\frac{\kappa ^4
\epsilon _{\kappa}^{2}}{4\pi ^2\sqrt{12}}e^{-\sqrt{12}\epsilon _{\kappa}}~.
\eqno(A11)
$$

\bigskip
\noindent
5. {\bf The catenary}.

\noindent
$x^{\mu}(s)=\sigma ^{-2}(\kappa {\rm sinh}\sigma s,\kappa {\rm cosh}\sigma
s, \tau\sigma s,0)$ and $x^{\mu}(0)=\sigma ^{-2} (0,\kappa,0,0)$. The integral
is of the type
$$
{\rm I}_5=
\int _{-\infty}^{+\infty}\frac{\sigma ^{3}e^{-i\epsilon _{\sigma}u} du}{2\kappa ^2({\rm cosh} u-1)-
\tau ^2u^2}~.  
\eqno(A12)
$$
This integral turns into $I_2$ and $I_4$ in the limits mentioned in the text,
respectively, but again there is no helpful formula in the GR {\em Table}, and
thus ${\rm I}_5$ appears to be non-analytic.

\bigskip
\noindent
6. {\bf The helicoid (helix of variable pitch)}.

\noindent
$x^{\mu}(s)=(\frac{\Delta}{RR_{+}}{\rm sinh} (R_{+}s), \frac{\Delta}{RR_{+}}
{\rm cosh}(R_{+}s), \frac{\kappa\tau}{R\Delta R_{-}}{\rm cos}(R_{-}s),
\frac{\kappa \tau}{R\Delta R_{-}}{\rm sin}(R_{-}s))$ and
$x^{\mu}(0)=(0,\frac{\Delta}{RR_{+}},\frac{\kappa \tau}{R\Delta R_{-}},0)$,
where $\Delta ^2=\frac{1}{2}(R^2+\kappa ^2 +\tau ^2 +\nu ^2)$;
$R^4=(\kappa ^2+\tau ^2 +\nu ^2)^2-4\kappa ^2 \tau ^2$;
$R_{+}^2=\frac{1}{2}(R^2+\kappa ^2 -\tau ^2 -\nu ^2)$;
$R_{-}^2=\frac{1}{2}(R^2-\kappa ^2 +\tau ^2 +\nu ^2)$;
leading to the following integral
$$
{\rm I}_6=
\frac{1}{2}\int _{-\infty}^{+\infty}
\frac{e^{-iEs}ds}{(\frac{\Delta}{RR_{+}})^2[{\rm cosh} (R_{+}s)
-1]+(\frac{\kappa \tau}{R\Delta R_{-}})^2[{\rm cos}(R_{-}s)-1]}~.
\eqno(A13)
$$
This is the most complicated non-analytic stationary case, with no helpful
formula in the GR {\em Table}.

\nonumsection{References}


\end{document}